\documentclass[twocolumn]{aa}
\usepackage{graphicx}
\usepackage{txfonts}
\usepackage{natbib}
\bibpunct{(}{)}{;}{a}{}{,} 
\newcommand{\igr}  {IGR J21343+4738}
\newcommand{\ha}  {H$\alpha$}
\newcommand{\ew}  {EW(H$\alpha$)}

\def\simless{\mathbin{\lower 3pt\hbox
     {$\rlap{\raise 5pt\hbox{$\char'074$}}\mathchar"7218$}}}   
\def\simmore{\mathbin{\lower 3pt\hbox
     {$\rlap{\raise 5pt\hbox{$\char'076$}}\mathchar"7218$}}}   

\def\msun{~{\rm M}_\odot}

\begin{document}

   \title{Disc-loss episode in the Be shell optical counterpart to the high
   mass X-ray binary IGR J21343+4738}

   \subtitle{}
  \author{
	P. Reig\inst{1,2}
	\and
  	A. Zezas\inst{2,3}
          }

\authorrunning{Reig et al.}
\titlerunning{The optical counterpart to IGR J21343+4738}

   \offprints{pau@physics.uoc.gr}

   \institute{IESL, Foundation for Reseach and Technology-Hellas, 71110, 
   		Heraklion, Greece 
	 \and Physics Department, University of Crete, 71003, 
   		Heraklion, Greece 
		\email{pau@physics.uoc.gr}
	 \and Harvard-Smithsonian Center for Astrophysics, 60 Garden
	 Street, Cambridge, MA02138, USA
	}

   \date{Received ; accepted}

\abstract
{
Present X-ray missions are regularly discovering new  X/$\gamma$-ray
sources. The identification of their counterparts at other wavelengths allows 
us to determine their nature. 
}
{The main goal of this work is to determine the properties of the optical
counterpart to the INTEGRAL source IGR J21343+4738, and study its long-term
optical variability. Although its nature as a Be/X-ray binary has been 
suggested, little is known about its physical parameters.}
{
We have been monitoring IGR J21343+4738 since 2009 in the optical band. 
We present optical photometric $BVRI$ and spectroscopic observations
covering the wavelength band 4000-7500 \AA. The photometric data
allowed us to derive the colour excess $E(B-V)$ and estimate the distance.
The blue-end spectra were used to determine the spectral type of the optical 
companion, while the spectra around the H$\alpha$ line allowed us to study the 
long-term structural changes in the circumstellar disc.  
}
{
We find that the optical counterpart to IGR J21343+4738 is a $V=14.1$
B1IVe shell star located at a distance of $\sim$8.5 kpc.  The H$\alpha$ line
changed from an absorption  dominated  profile to an emission dominated
profile, and then back again into absorption. In addition, fast V/R 
asymmetries developed once the disc develops. 
Although the  Balmer lines
are the most strongly affected by shell absorption, we find that shell
characteristics are also observed in He I lines. 
 }
{The optical spectral variability of IGR J21343+4738 is attributed to the
formation of an equatorial disc around the Be star and the development of
an enhanced density perturbation that revolves inside the disc. We have
witnessed the formation and dissipation of the circumstellar disc.  The
strong shell profile of the \ha\ and He I lines and the fact that no 
transition from shell phase to a pure emission phase is seen imply that 
we are seeing the system near edge-on. }


   \maketitle

\begin{table*}
\caption{Photometric measurements of the optical counterpart to \igr.}
\label{phot}
\begin{center}
\begin{tabular}{l c c c c c}
\hline	\hline
Date &  JD (2,400,000+)    &   $B$  &   $V$   &   $R$  & $I$   \\
\hline
30-06-2009 &55013.44	&14.76$\pm$0.02  &14.21$\pm$0.02  &13.86$\pm$0.02  &13.48$\pm$0.03    \\
26-08-2011 &55800.50	&14.80$\pm$0.02  &14.24$\pm$0.01  &13.89$\pm$0.01  &13.48$\pm$0.03    \\
09-09-2011 &55814.31	&14.77$\pm$0.02	 &14.22$\pm$0.02  &13.88$\pm$0.01  &13.46$\pm$0.03    \\
29-07-2013 &56503.36	&14.57$\pm$0.02  &14.06$\pm$0.02  &13.72$\pm$0.02  &13.36$\pm$0.02    \\
29-08-2013 &56534.45	&14.62$\pm$0.02  &14.12$\pm$0.02  &13.76$\pm$0.02  &13.39$\pm$0.04    \\
\hline
\end{tabular}
\end{center}
\end{table*}
\begin{table*}
\caption{Log of the spectroscopic observations around the \ha\ line and results from the
spectral analysis.}
\label{red}
\centering
\begin{tabular}{@{~~}l@{~~}c@{~~}l@{~~}l@{~~}c@{~~}c@{~~}c@{~~}c@{~~}c}
\noalign{\smallskip}	\hline \noalign{\smallskip}
Date	&JD 		&Telescope  	&Wavelength	 &\ew        &$\log(V/R)$	&$F_{\rm V}/F_{\rm cd}$	&$F_{\rm R}/F_{\rm cd}$ &Velocity\tablefootmark{*}\\
	&(2,400,000+)	&		&coverage (\AA)	 &($\AA$)    & 			&			& 			&shift (km s$^{-1}$)     \\
\noalign{\smallskip}\hline\noalign{\smallskip}
30-07-2009  		&55043.51	&SKO	&5290--7365	&$+1.3\pm0.3$ 	&$+0.14\pm0.10$	&1.3	&1.3	&$ -52\pm17$ \\
29-09-2009  		&55104.37	&SKO    &6070--6980     &$+0.5\pm0.3$	&$-0.17\pm0.12$ &1.8    &1.9 	&$ -95\pm26$ \\
28-08-2010  		&55437.41	&SKO    &5305--7380     &$+0.4\pm0.2$   &$+0.07\pm0.06$ &1.9    &1.8	&$-102\pm18$ \\
30-09-2010\tablefootmark{\dag}  &55470.40&SKO &5220--7295     &$+0.4\pm0.2$   &$+0.06\pm0.06$ &1.6    &1.5	&$-127\pm43$ \\
20-08-2011  		&55794.45	&SKO    &5210--7285     &$-1.3\pm0.2$   &$-0.08\pm0.03$ &1.8    &1.9	&$ -85\pm16$ \\
06-09-2011\tablefootmark{\dag\dag}&55811.30&SKO&6080--6995    &$-1.8\pm0.4$	&$-0.12\pm0.07$ &2.0    &2.1	&$-117\pm9$ \\
24-08-2012  		&56164.41	&SKO    &5085--7160     &$-6.9\pm0.5$	&$+0.36\pm0.03$ &2.4    &1.7	&$ -62\pm20$ \\
06-09-2012  		&55177.46	&SKO    &5225--7600     &$-8.1\pm0.3$	&$+0.34\pm0.03$ &2.6    &1.9	&$ -61\pm6$ \\
13-09-2012  		&56184.40	&SKO    &5415--7490     &$-7.1\pm0.3$	&$+0.39\pm0.02$ &1.9    &1.4	&$ -36\pm38$ \\
15-10-2012  		&56216.81	&FLWO	&6200--7200	&$-5.3\pm0.2$	&$+0.24\pm0.03$	&3.1	&2.5 	&$-128\pm16$ \\
19-10-2012  		&56220.32	&SKO	&4745--6825	&$-5.4\pm0.3$	&$+0.12\pm0.02$	&1.9	&1.7	&$-137\pm24$ \\
26-12-2012\tablefootmark{\dag}  	&56288.36	&WHT	&6370--7280	&$-2.4\pm0.2$	&$-0.73\pm0.02$	&2.1	&3.3 	&$-131\pm4$ \\
04-01-2013		&56297.57	&FLWO	&6200--7200	&$-2.1\pm0.2$	&$-0.91\pm0.07$	&2.7	&4.7	&$-139\pm19$ \\
11-01-2013		&56304.58	&FLWO	&6200--7200	&$-1.1\pm0.2$	&$-0.77\pm0.09$	&2.0	&3.2	&$-150\pm11$ \\
15-06-2013		&56459.54	&SKO	&4950--7030	&$+2.1\pm0.3$	&--		&--	&-- 	&$-148\pm32$ \\
05-07-2013		&56477.86	&FLWO	&6200--7200	&$+1.4\pm0.2$	&--		&--	&-- 	&$-126\pm12$ \\
31-07-2013		&56505.45	&SKO	&5400--7480	&$+1.7\pm0.2$	&--		&--	&-- 	&$ -69\pm11$ \\
23-08-2013\tablefootmark{\dag}	&56528.59	&WHT	&6350--7235	&$+1.7\pm0.1$	&--		&--	&-- 	&$-105\pm4$ \\
\noalign{\smallskip}	\hline
\end{tabular}
\tablefoot{
\tablefoottext{$\dag$}{Average of three spectra.} 
\tablefoottext{$\dag\dag$}{Average of two spectra.} 
\tablefoottext{*}{Heliocentric scale.}
}
\end{table*}
   
\section{Introduction}

The first report of a detection of \igr\ in the hard X-ray band is found in
the {\it INTEGRAL}/IBIS all-sky survey catalogues, which are based on data
taken before the end of 2006 \citep{krivonos07,bird07}. At that time, its
nature was unknown, other than it was a transient source, detected during a
series of observations between December 2002 and February 2004 but which
was below the threshold of the {\it INTEGRAL} detectors between 2004-2007
\citep{bikmaev08}. A {\it Chandra} observation was performed on 18 December
2006, that is, during the off state of the {\it INTEGRAL} instruments.
However, a weak source consistent with the position of \igr\ was detected.
The {\it Chandra} observation allowed the refinement of its X-ray position
and the suggestion of an optical counterpart \citep{sazonov08}. 
Low-resolution ($FWHM \sim 15$ \AA) optical spectroscopic observations of
the likely counterpart indicated a B3 star. Although the
H$\alpha$ line was found in absorption, \igr\ was proposed to be a
high-mass X-ray binary with a Be star companion. It was argued that the
star was going through a disc-loss episode at the time of the observations
\citep{bikmaev08}.  

Be/X-ray binaries are a class of high-mass X-ray binaries that consist of a
Be star and a neutron star \citep{reig11}.  The mass donor in these systems
is a relatively massive ($\simmore 10 \msun$) and fast-rotating
($\simmore$80\% of break-up velocity) star, whose equator is surrounded by
a disc formed from photospheric plasma ejected by the star.  \ha\ in
emission  is typically the dominant feature in the spectra of such stars.
In fact, the strength of the Balmer lines in general and of \ha\ in
particular (whether it has ever been in emission) together with a
luminosity class III-V constitute the defining properties of this class of
objects.  The equatorial discs are believed to be quasi-Keplerian and
supported by viscosity \citep{okazaki01}.  The shape and strength of the
spectral emission lines are useful indicators of the state of the disc.
Global disc variations include the transition from a Be phase, i.e., when
the disc is present, to a normal B star phase, i.e., when the disc is
absent and also cyclic V/R changes, i.e., variation in the ratio of the
blue to red peaks of a split profile that are attributed to the precession
of a density perturbation inside the disc \citep{okazaki91,papaloizou06}.

In this work we present the first long-term study of the optical
counterpart to the X-ray source \igr\ and report a disc-loss episode. The
absence of the disc allows us to derive some of the fundamental physical
parameters such as reddening, distance, and rotation velocity, without
contamination from the disc. We also confirm that \igr\ is a Be/X-ray
binary, although with an earlier spectral type than the one suggested by
\citet{bikmaev08}.

\begin{table}
\caption{Log of the spectroscopic observations in the blue region.}
\label{blue}
\centering
\begin{tabular}{@{~~}l@{~~}c@{~~}l@{~~}l@{~~}c}
\noalign{\smallskip}	\hline \noalign{\smallskip}
Date	&JD 			&Telescope  	&Wavelength	&num. of	       \\
	&(2,400,000+)		&		&coverage (\AA)	&spectra	       \\
\noalign{\smallskip}\hline\noalign{\smallskip}
13-09-2012  	&56184.34	&SKO    &3810--5165	&3 \\
14-09-2012	&56185.31	&SKO    &3800--5170	&5 \\
15-10-2012  	&56216.74	&FLWO	&3890--4900	&5 \\
26-12-2012  	&56288.36	&WHT	&3870--4670	&3 \\
15-07-2013	&56488.87	&FLWO	&3900--4900	&3 \\
23-08-2013	&56528.58	&WHT	&3840--4670	&3 \\
\noalign{\smallskip}	\hline
\end{tabular}
\end{table}

\begin{figure*}
\resizebox{\hsize}{!}{\includegraphics{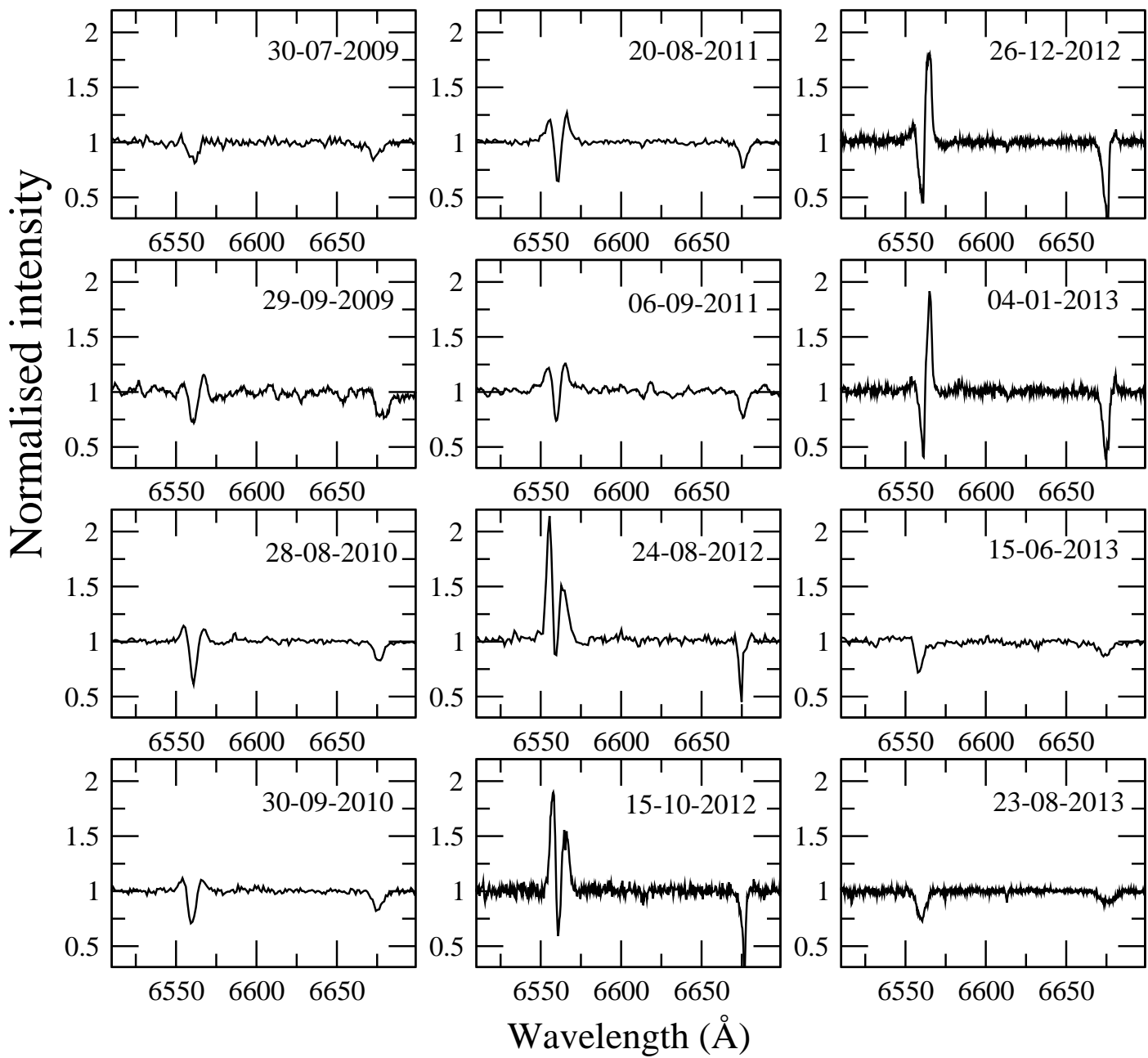}} 
\caption[]{Evolution of the \ha\ and He I 6678 \AA\ lines. Absorption by
the disc along the line of sight produces very narrow lines (shell 
profiles). }
\label{haprof}
\end{figure*}


\section{Observations}

Optical spectroscopic and photometric observations of the optical
counterpart to the INTEGRAL source \igr\ were obtained from the 1.3m
telescope of the Skinakas observatory (SKO) in Crete (Greece) and from the
Fred Lawrence Whipple Observatory (FLWO) at Mt. Hopkins (Arizona).  In
addition, \igr\ was observed in service time with the 4.2-m William
Herschel Telescope telescope (WHT) of El Roque de los Muchachos observatory
in La Palma (Spain).

The 1.3\,m telescope of the Skinakas Observatory was equipped with a
2000$\times$800 ISA SITe CCD and a 1302 l~mm$^{-1}$ grating, giving a
nominal dispersion of $\sim$1.04 \AA/pixel. On the nights 29 September 2009
and 6 September 2011, a 2400 l~mm$^{-1}$ grating with a dispersion of
$\sim$0.46 \AA/pixel was used. We also observed \igr\ in queue mode with
the 1.5-m telescope at Mt. Hopkins (Arizona), and the FAST-II spectrograph
\citep{fabricant98} plus FAST3 CCD, a backside-illuminated 2688x512 UA
STA520A chip with 15$\mu$m pixels and a 1200 l~mm$^{-1}$ grating (0.38
\AA/pixel). The WHT spectra were obtained in service mode on the nights 26
December 2012 and 23 August 2013 with the ISIS spectrograph and the R1200B
grating plus the EEV12 4096$\times$2048 13.5-$\mu$m pixel CCD (0.22
\AA/pixel) for the blue arm and the R1200R grating and REDPLUS
4096$\times$2048 15-$\mu$m pixel CCD (0.25 \AA/pixel) for the red arm.  The
spectra were reduced with the dedicated packages for spectroscopy of the
{\tt STARLINK} and  {\tt IRAF} projects following the standard procedure.
In particular, the FAST spectra were reduced with the FAST pipeline
\citep{tokarz97}. The images were bias subtracted and flat-field corrected.
Spectra of comparison lamps were taken before each exposure in order to
account for small variations of the wavelength calibration during the
night. Finally, the spectra were extracted from an aperture encompassing
more than 90\% of the flux of the object. Sky subtraction was performed by
measuring the sky spectrum from an adjacent object-free region. To ensure
the homogeneous processing of the spectra, they were normalized with
respect to the local continuum, which  was rectified to unity by employing
a spline fit.

The photometric observations were made from the 1.3-m telescope of the
Skinakas Observatory. \igr\ was observed through the Johnson/Bessel $B$,
$V$, $R$, and $I$ filters \citep{bessel90}. For the photometric
observations the telescope was equipped with a  2048$\times$2048 ANDOR CCD
with a 13.5 $\mu$m pixel size (corresponding to 0.28 arcsec on the sky) and
thus provides a field of view of 9.5 arcmin $\times$ 9.5 arcmin. The gain
and read out noise of the CCD camera at a read-out velocity of 2
$\mu$s/pixel are 2.7 $e^{-}$/ADU and 8 $e^{-}$, respectively. The FWHM
(seeing estimate) of the point sources in the images varied from 4 to 6
pixels (1.1''--1.7'') during the different campaigns.    Reduction of the
data was carried out in the standard way using the IRAF tools for aperture
photometry. First, all images were bias-frame subtracted and flat-field
corrected using twilight sky flats to correct for pixel-to-pixel variations
on the chip. The resulting images are therefore free from the instrumental
effects. All the light inside an aperture with radius 4.5'' was summed up
to produce the instrumental magnitudes. The sky background was determined
as the statistical mode of the counts inside an annulus 5 pixels wide and
20 pixels from the center of the object. The absorption caused by the
Earth's atmosphere was taken into account by nightly extinction corrections
determined from measurements of selected stars that also served as
standards. Finally, the photometry was accurately corrected for colour
equations and transformed to the standard system using nightly observations
of standard stars from Landolt's catalogue \citep{landolt92,landolt09}. The
error of the photometry was calculated as the root-mean-square of the
difference between the derived final calibrated magnitudes of the standard
stars and the magnitudes of the catalogue.

The photometric magnitudes are given in Table~\ref{phot}, while information
about the spectroscopic observations can be found in Tables~\ref{red} and
\ref{blue}.

\begin{figure}
\resizebox{\hsize}{!}{\includegraphics{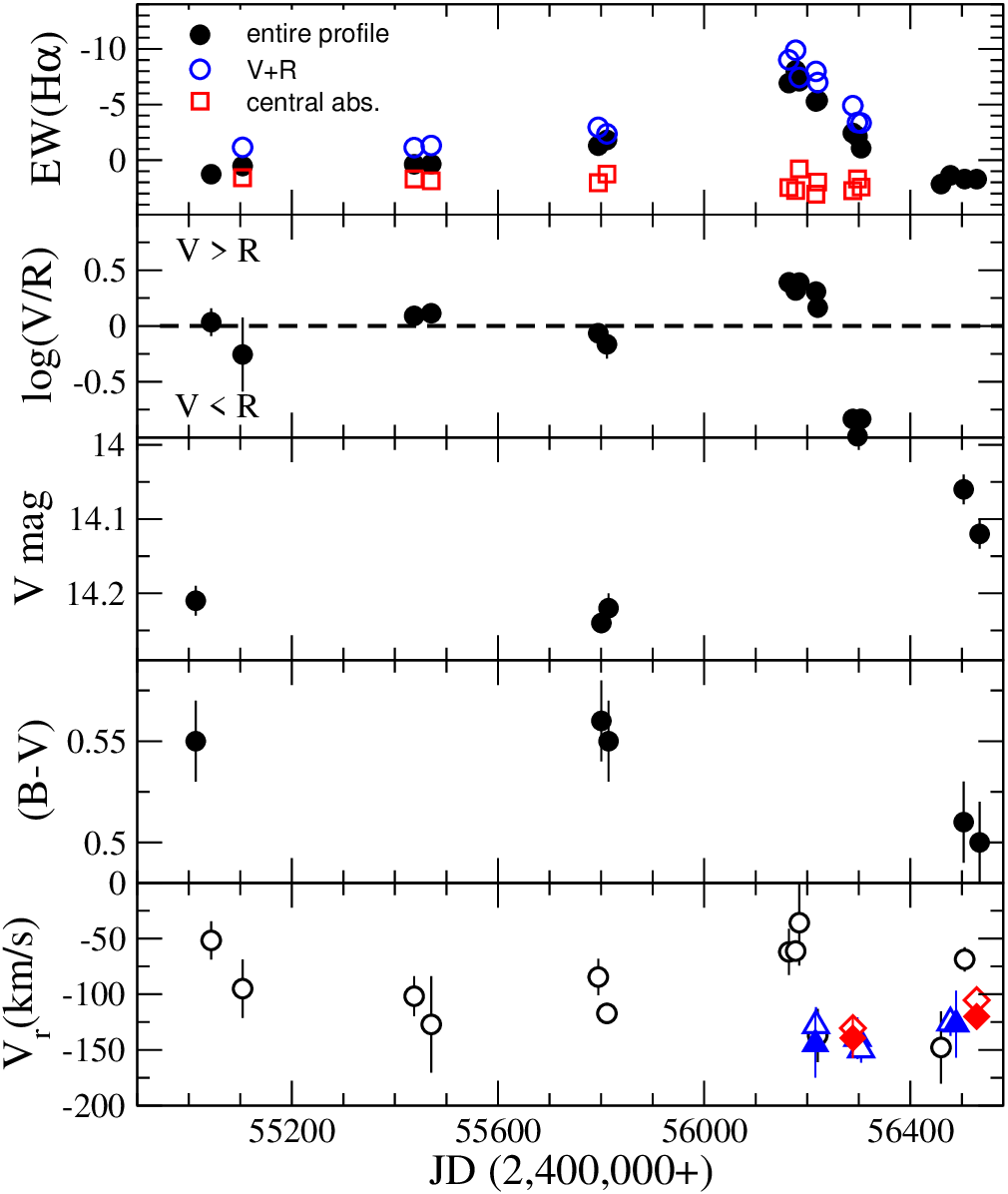}} 
\caption[]{From top to bottom, evolution of the \ha\ equivalent width, 
V/R ratio, $V$ magnitude, $(B-V)$ colour, and velocity shift with time.}
\label{specpar}
\end{figure}

\begin{figure*}
\begin{center}
\includegraphics[width=16cm,height=10cm]{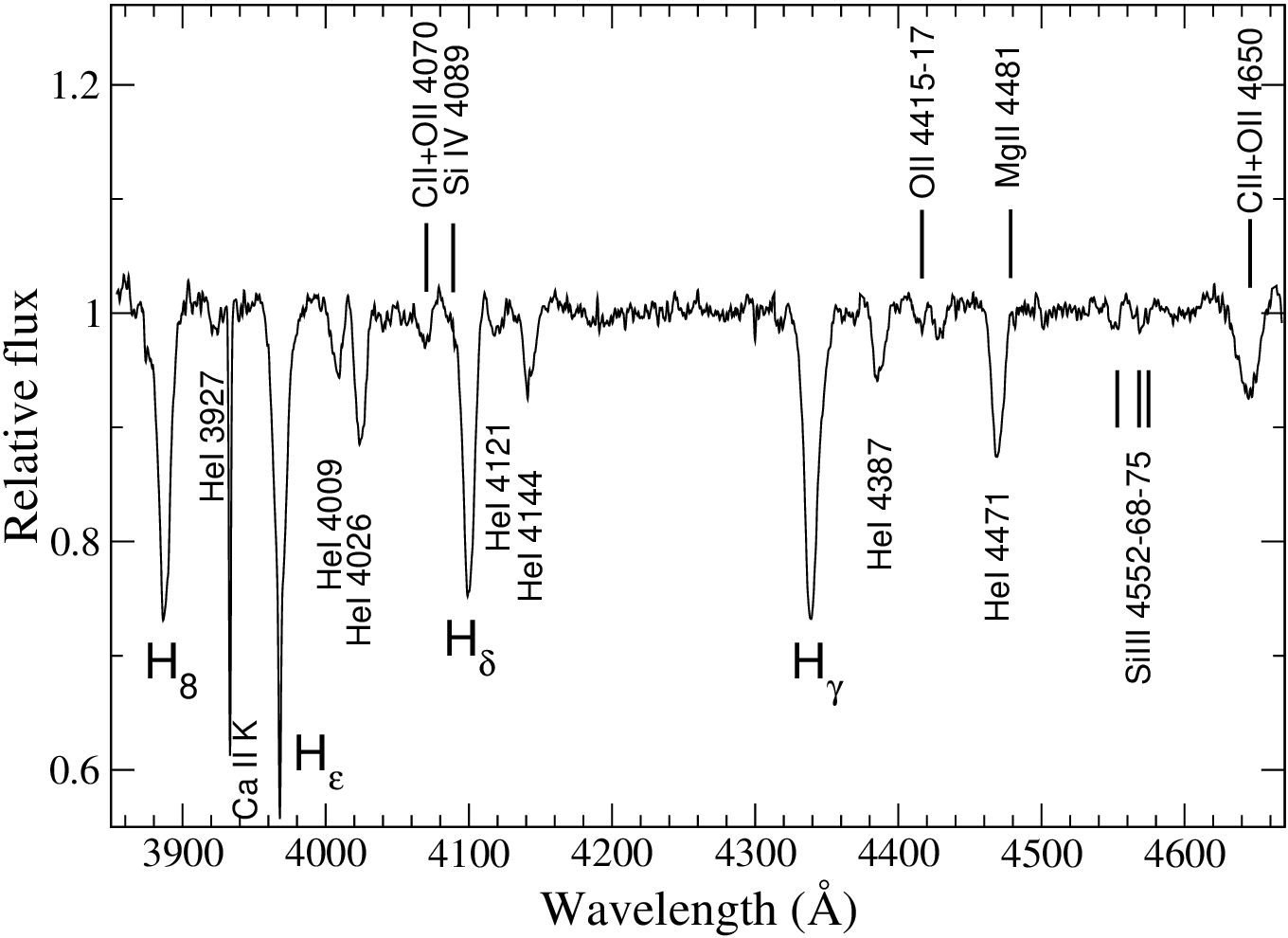} 
\caption[]{WHT spectrum of \igr\ and identified lines used for spectral
classification. The spectrum was smoothed with a Gaussian filter (FWHM=1).}
\label{wht}
\end{center}
\end{figure*}

\section{Results}

\subsection{The \ha\ line: evolution of spectral parameters}
\label{haevol}

The \ha\ line is the prime indicator of the circumstellar disc state. In
particular, its equivalent width (\ew)  provides a good measure of the size
of the circumstellar disc \citep{quirrenbach97,tycner05,grundstrom06}. \ha\
emission results from recombination of photoionised electrons by the
optical and UV radiation from the central star. Thus, in the absence of the
disc, no emission should be observed and the line should display an
absorption profile. The \ha\ line of the massive companion in \igr\ is
highly variable, both in strength and shape. When the line appears in
emission it always shows a double-peaked profile, but the relative
intensity of the blue (V) over the red (R) peaks varies. The central
absorption that separates the two peaks goes beyond the continuum, placing 
\igr\ in the group of the so-called {\em shell} stars
\citep{hanuschik95,hanuschik96a,hummel00,rivinius06}. Figure \ref{haprof}
displays the evolution of the line profiles. V/R variability is clearly
seen, indicating a distorted disc \citep{hummel97}. Significant changes in
the structure of the equatorial disc on timescales of months are observed.
In addition, a long-term growth/dissipation of the disc is suggested by the
increase of the equivalent width and subsequent decrease.

Table~\ref{red} gives the log of the spectroscopic observations and some
important parameters that resulted from fitting  Gaussian functions to the
\ha\  line profile. Due to the deep central absorption, three Gaussian
components (two in emission and one in absorption) were generally needed to
obtain good fits. Column 5 gives the equivalent width of the entire (all
components) \ha\ line. The main source of uncertainty in
the equivalent width stems from the always difficult definition of the
continuum. The \ew\ given in Table~\ref{red} correspond to the average of
twelve measurements, each one from a different definition of the continuum
and the quoted error is the scatter (standard deviation) present in those
twelve measurements.

Column 6 shows the ratio between the core intensity of the blue and red
humps. The V/R ratio is computed as the logarithm of the ratio of the
relative fluxes  at the blue and red emission peak maxima. Thus negative
values indicate a red-dominated peak, that is, $V<R$, and positive values a
blue-dominated line, $V>R$ .

Columns 7 and 8 in Table~\ref{red} give the ratio of the peak flux
of each component over the minimum flux of the deep absorption core. This
ratio simply serves to confirm the shell nature of \igr\ in a more
quantitative way. \citet{hanuschik96a} established an empirical {\em shell
criterion} based on the H$\alpha$ line , whereby shell stars are those with
$F_{\rm p}/F_{\rm cd}\simmore 1.5$, where $F_{\rm p}$ and $F_{\rm cd}$ are
the mean peak and trough flux, respectively.

Column 9 is the velocity shift of the central narrow absorption shell
feature when the disc is present or of the absorption profile of the \ha\
line in the absence of the disc. Prior to the measurement of the velocity
shift, all the spectra were aligned taking the value of the insterstellar
line at 6612.8 \AA\ as reference.  Note that these shifts do not
necessarily represent the radial velocity of the binary, as the \ha\ line
is strongly affected by circumstellar matter \citep[see e.g.,][for a
discussion of the various effects when circumstellar matter is present in
the system]{harmanec03}. Typically, the He I lines in the blue end part of
the spectrum are used for radial velocity studies. Note, however, that in
\igr, even these lines are affected by disc emission (see
Sect.~\ref{diskc}). Nevertheless, we measured the radial velocity of the
binary by cross-correlating the higher resolution blue-end spectra obtained
from the FLWO and WHT (Table~\ref{blue}) with a template using the {\em
fxcor} task in the {\it IRAF} package. This template was generated from the
BSTAR2006 grid of synthetic spectra \citep{lanz07} and correspond to a
model atmosphere with $T_{\rm eff}=25000$ K, $log g=3.75$ convolved by a
rotational profile with $v \sin i=380$ km s$^{-1}$. The results for the 
July and August 2013 observations, when the contribution of the disc is
expected to be minimum, are $v_r=-127\pm30$ km s$^{-1}$ (HJD 2,456,488.863)
and $v_r=-120\pm10$ km s$^{-1}$ (HJD 2,456,528.582), respectively.

Figure \ref{specpar} shows the evolution of \ew, the V/R ratio, the V
magnitude, the $(B-V)$ colour, and the velocity shift with time. In the
top panel of this figure, different symbols represent the equivalent width
of the different components of the line. Open circles give the sum of the
equivalent widths of the individual  V and R peaks, while the squares are
the equivalent width of the deep central absorption. These values were
obtained from the Gaussian fits. The overall equivalent width (filled
circles) was measured directly from the spectra.  In the bottom panel
of Fig.~\ref{specpar}, open symbols correspond to the velocity shifts
measured from the \ha\ line, while filled symbols are radial velocities
obtained by crossc-orrelating the 3950--4500 \AA\ spectra with the
template.  Black circles denote SKO spectra, blue triangles correspond to
data taken from the FLWO, and red diamonds spectra obtained with the
WHT.

\subsection{Spectral classification}
\label{specl}

Figure~\ref{wht} shows the average blue spectrum of \igr\ obtained with the
4.2-m WHT on the night 23 August 2013. The main spectral features have been
identified. The 3900--4600 \AA\ spectrum  is dominated by hydrogen and
neutral helium absorption lines, clearly indicating an early-type B star.
The earliest classes (B0 and B0.5) can be ruled out because no ionised
helium is present. However, \ion{Si}{III} 4552-68-75 is crearly detected,
favouring a spectral type B1-B1.5. The relatively weakness of of
\ion{Mg}{II} at 4481 \AA\ also indicates a spectral type earlier than B2. 
The strength of the \ion{C}{III}+\ion{O}{II} blend at 4070 \AA\ and 4650
\AA\ agrees with this range (B1--B2) and points toward an evolved star. A
subgiant or giant star, i.e., luminosity class IV or III, is also favoured
by the presence of \ion{O}{II} at 4415-17 \AA. However, in this case, the
triplet {\ion{Si}{III} 4552-68-75 \AA\ and \ion{Si}{IV} 4089 \AA\ should be
stronger than observed. We conclude that the spectral type of the optical
counterpart to \igr\ is in the range B1--B1.5 V--III, with a prefered
classification of B1IV.

\begin{figure}
\resizebox{\hsize}{!}{\includegraphics{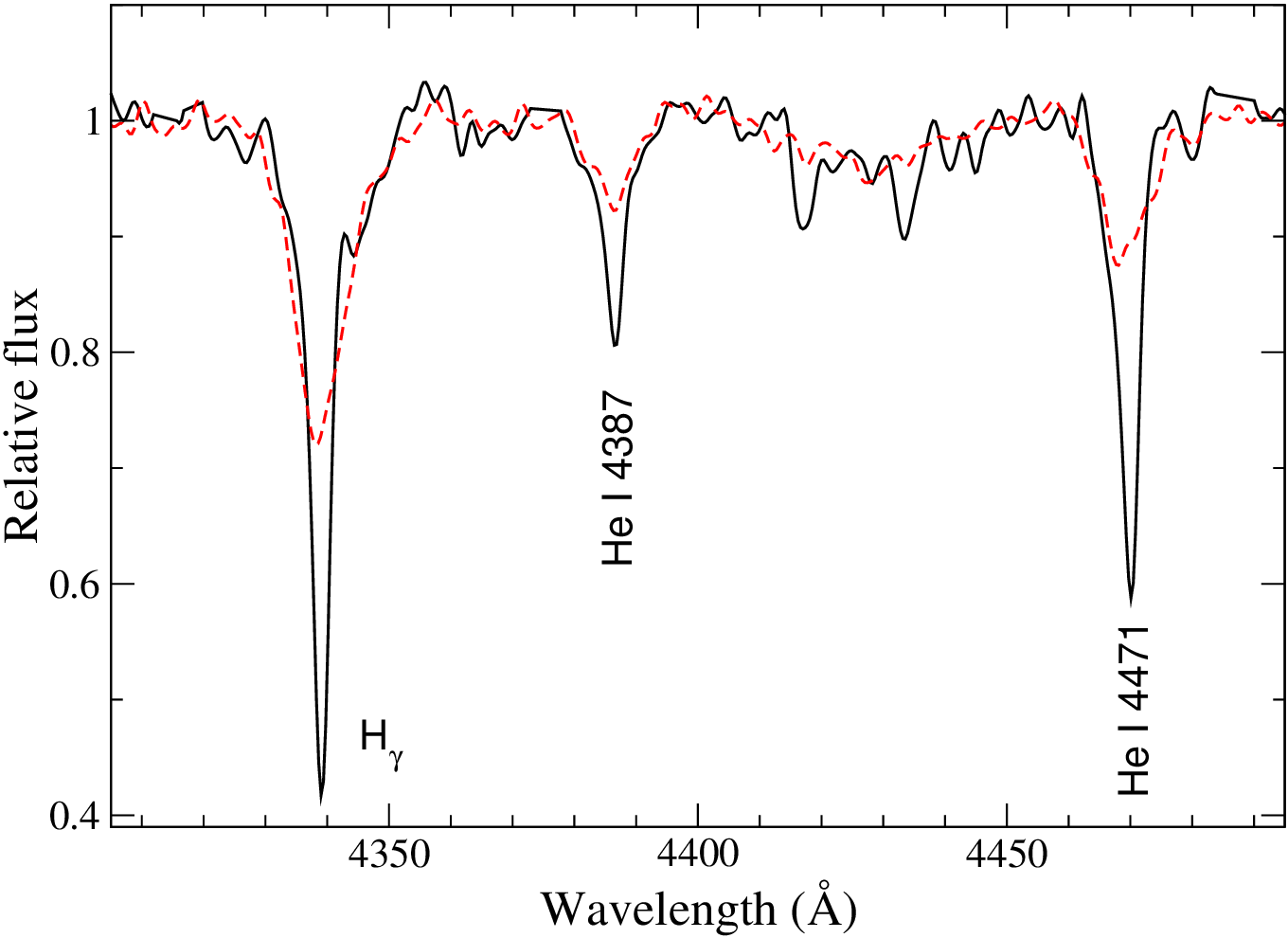}} 
\caption[]{Comparison of two spectra of \igr\ at different epochs, when the
disc was present (solid black line) and when the disc presumably had
vanished (dashed red line). The spectra were taken from the FLWO on 15
October 2012 and 15 July 2013, respectively.}
\label{widthcomp}
\end{figure}

\subsection{Contribution of the disc to the spectral lines and colours}
\label{diskc}

Figure~\ref{widthcomp} shows a comparison of two blue-end spectra of \igr\
taken from the FLWO at two different epochs. The October 2012 spectrum
corresponds to a Be phase when the \ha\ line was strongly in emission,
while the July 2013 spectrum corresponds to a B phase when this line
displayed an absorption profile. One of the most striking results that can
be directly derived from the visual comparison of the spectra is the
significantly narrower width of the spectral lines, particularly those of
the Balmer series and the He I lines, when the disc is present. To estimate
the contribution of the disc to the width of the lines we measured the FWHM
of the  \ion{He}{I} at 6678 \AA\ (Fig.~\ref{haprof}) as a function of time.
The result can be seen in Fig.~\ref{hei}, where the evolution of \ew\ with
time is also plotted.  The width of the helium line was significantly
narrower and the core deeper during the strong shell phase (spectra taken
during 2012, MJD $\sim$56100--56200) than at instances where the disc was
weak, in 2009 and 2013). 

The disc emission also affects the photometric magnitudes and colours. The
observed $(B-V)$ colour when the disc was present (observation taken in
2011) is 0.05 mag larger than during 2013 when the disc disappeared. That
is, the disc introduces and extra reddening component.

\begin{figure}
\resizebox{\hsize}{!}{\includegraphics{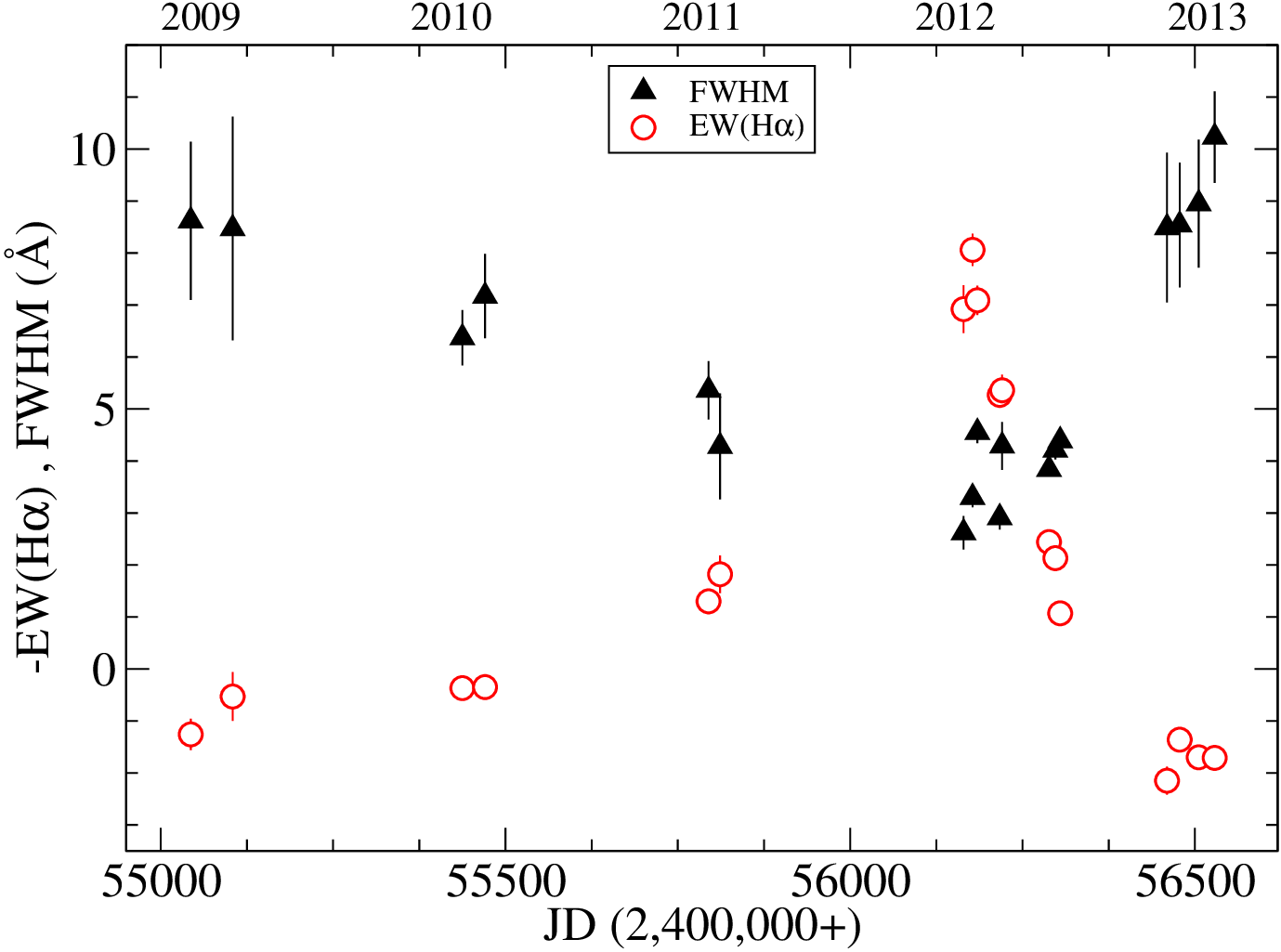}} 
\caption[]{Evolution of the full width half maximum of the He I
$\lambda$6678 line. Note the progressive narrowing of the line as the shell
phase develops and the large values when \ew\ is positive (interpreted as 
the absence of the disc).}
\label{hei}
\end{figure}

\subsection{The He I lines: rotational velocity}
\label{rotvel}

Shell stars are Be stars with strongly rotationally broadened photospheric
emission lines with a deep absorption core \citep{rivinius06}.  The
rotational velocity is believed to be a crucial parameter in the formation
of the circumstellar disc. A rotational velocity close to the break-up or
critical velocity (i.e. the velocity at which centrifugal forces balance 
Newtonian gravity) reduces the effective equatorial gravity to the extent
that weak processes such as gas pressure and/or non-radial pulsations may
trigger the ejection of photospheric matter with sufficient energy and
angular momentum to make it spin up into a Keplerian disc. Because stellar
absorption lines in Be stars are rotationally broadened, their widths can
be used to estimate the projected rotational velocity, $v \sin i$, where
$v$ is the equatorial rotational velocity and $i$ the inclination angle
toward the observer.  However, to obtain a reliable measurement of the
rotational velocity of the Be star companion the He I lines have to be free
of disc emission. As shown in the previous section, the width of the
line becomes narrower as the disc grows, underestimating the value of the
rotational velocity.

We estimated the projected rotational velocity of \igr\ by measuring
the full width at half maximum (FWHM) of He I lines, following the
calibration by \citet{steele99}. These authors used four neutral helium
lines, namely 4026 \AA, 4143 \AA, 4387 \AA, and 4471 \AA, to derive
rotational velocities. We measured the width of these lines from the August
2013 WHT spectrum as it provides the highest resolution in our sample and
correspond to a disc-loss phase, as indicated by the absorption profile of
the \ha\ line. We made five different selections of the continuum and
fitted Gaussian profiles to these lines.  We also corrected the lines for
instrumental broadening by subtracting in quadrature the FWHM of a nearby
line from the calibrated spectra.  The projected rotational velocity
obtained as the average of the values from the four He I lines was
$v \sin i=365\pm15$ km s$^{-1}$. The quoted errors are the standard
deviation of all the measurements. 

The rotational velocity can also be estimated by comparing the
high-resolution August 2013 WHT spectrum with a grid of synthetic spectra
broadened at various values of the rotational velocity.  We employed the
BSTAR2006 grid \citep{lanz07}, which uses the code TLUSTY
\citep{hubeny88,hubeny92,hubeny94} to create the model atmosphere and
SYNSPEC\footnote{http://nova.astro.umd.edu} to calculate the emergent
spectrum. We assumed a model atmosphere with solar composition, $T_{\rm
eff}=25000$ K and $\log g=3.50$ and a microturbulent velocity of 2 km
s$^{-1}$. This spectrum was convolved with rotational and intrumental
(gaussian) profiles using ROTIN3.  Thirteen rotational velocities from 300
km s$^{-1}$ to 420 km s$^{-1}$ with steps of 10 km s$^{-1}$ were
considered. The rotational velocity that minimises the sum of the squares
of the difference between data and model corresponded to $v \sin i=380$ km
s$^{-1}$, consistent with the previous value.

\subsection{Reddening and distance}

To estimate the distance, the amount of reddening to the source has to be
determined. In a Be star,  the total measured reddening is made up of two
components: one produced mainly by dust in the interstellar space through
the line of sight and another produced by the circumstellar gas around the
Be star \citep{dachs88,fabregat90}. Although the  physical origin and
wavelength dependence of these two reddenings is completely different,
their final effect upon the colours is very difficult to disentangle
\citep{torrejon07}. In fact, interstellar reddening is caused by {\em
absorption} and {\em scattering} processes, while circumstellar reddening
is due to extra {\em emission} at longer wavelenths. The disc-loss episode
observed in \igr\ allows us to derive the true magnitudes and colours of
the underlying Be star, without the contribution of the disc. Thus the
total reddening measured during a disc-loss episode corresponds entirely to
interstellar extinction.

The observed colour of \igr\ in the absence of the disc is
$(B-V)=0.50\pm0.02$ (Table~\ref{phot}), while the expected one for a
B1--1.5V--IV star is $(B-V)_0=-0.26$
\citep{johnson66,fitzgerald70,gutierrez-moreno79,wegner94}. Thus we derive
a colour excess of $E(B-V)=0.76\pm0.02$ or visual extinction $A_{\rm V}=R
\times E(B-V)= 2.4\pm0.1$, where the standard extinction law $R=3.1$ was
assumed. Taking an average absolute magnitude  of $M_V=-3.0$, typical of a
star of this spectral type \citep{humphreys84,wegner06}, the distance to
\igr\ is estimated to be 8.7$\pm$1.3 kpc. The final error was obtained by
propagating the errors of $B-V$  (0.02 mag), $A_V$ (0.1 mag) and $M_V$ (0.3
mag).

\section{Discussion}
\label{discussion}

We have investigated the long-term variability of the optical counterpart
to the X-ray source \igr. \citet{bikmaev08} suggested that \igr\ is a
high-mass X-ray binary with a B3e companion, even though their
spectroscopic observations showed \ha\ in absorption. They argued that the
system could be in a disc-loss phase. We confirm the Be nature of \igr, but
suggest an earlier type companion, in agreement with the spectral type
distribution of Be/X-ray binaries in the Milky Way.  All spectroscopically
identified optical companions of  Be/X-ray binaries in the Galaxy do not
have spectral type later than B2 \citep{negueruela98b}. 

\begin{table*}
\caption{Comparison of the characteristic time scales of \igr\ with other 
Be/X-ray binaries. $P_{\rm orb}$ is the orbital period, $T_{\rm V/R}$is 
the time needed to complete a V/R cycle, and $T_{\rm disc}$ is the time 
for the formation and dissipation of the disc. }
\label{comp}
\centering
\begin{tabular}{@{~~}l@{~~}l@{~~}c@{~~}c@{~~}c@{~~}c@{~~}c@{~~}l}
\noalign{\smallskip} \hline \noalign{\smallskip}
X-ray		&Spectral	&Disc-loss      &P$_{\rm orb}$   &$T_{\rm V/R}$	&$T_{\rm disc}$	&Reference \\
source		&type		&episode\tablefootmark{$^\dag$}&(days)	&(year)	&(year)	& \\
\noalign{\smallskip} \hline \noalign{\smallskip}
\igr		&B1V-IV	 	&yes		&--		&0.5-0.8	&5--6	&This work \\
4U 0115+634	&B0.2V  	&yes		&24.3		&0.5--1 	&3--5	&1,2      \\
RX J0146.9+6121	&B1III-V	&no		&--		&3.4    	&--	&3    \\
V 0332+53	&O8-9V  	&no		&34.2		&1      	&--	&4    \\
X-Per		&O9.5III	&yes		&250		&0.6--2 	&7	&5,6,7  \\
RX J0440.9+4431	&B1III-V	&yes		&--		&1.5--2     	&$>10$	&8      \\
1A 0535+262	&O9.7III	&yes		&111		&1--1.5 	&4--5	&9,10,11 \\
IGR J06074+2205	&B0.5IV 	&yes		&--		&--      	&4--5	&12      \\
RX J0812.4-3114	&B0.5III-V	&yes		&81.3		&--     	&3--4	&13 \\
4U 1145-619	&B0.2III	&no		&187		&3      	&--	&14  \\
4U 1258-61	&B2V		&yes		&132		&0.36     	&--	&15  \\
SAX J2103.5+4545&B0V		&yes		&12.7		&--		&1.5--2	&16 \\
\noalign{\smallskip} \hline
\end{tabular}
\tablefoot{
\tablefoottext{$^\dag$}{By disc-loss episodes we mean periods when the \ew\ was seen to be positive.} 
}
\tablebib{
(1) \citet{negueruela01}; (2) \citet{reig07b}; (3) \citet{reig00} ; 
(4) \citet{negueruela98a}; (5) \citet{lyubimkov97} ;
(6) \citet{delgado01}; (7) \citet{clark01}; (8) \citet{reig05b} ;
(9) \citet{clark98}; (10) \citet{haigh04}; (11) \citet{grundstrom07} ;
(12) \citet{reig10b}; (13) \citet{reig01}; 
(14) \citet{stevens97} ; (15) \citet{corbetgx86} ; (16) \citet{reig10a}  
}
\end{table*}

\subsection{Spectral evolution and variability time scales}

Our monitoring of \igr\ reveals large amplitude changes in the shape and
strength of the spectral lines and two different time scales
associated with the variability of the disc: disc formation/dissipation is
estimated to occur on time scales of years, while V/R variability is seen
on time scales of months.

Our first observation was performed in July 2009 and shows the contribution
of a weak disc. Although the \ew\ is positive, indicating that absorption
dominates over emission, its value is smaller than that expected from a
pure photospheric line, which according to \citet{jaschek87} should be
$\sim$3.5--4 \AA. Also, the two peaks, V and R, separated by the central
depression can already be distinguished in our first spectrum.  The
strength of the \ha\ line increased and its shape changed from an
absorption dominated profile into an emission dominated one during the
period July 2009--September 2012. As the intensity increased, the line
became progressively more asymmetric. After September 2012, \ew\ began to
decrease with a faster rate than the rise. By Summer 2013, the \ha\ line
profile had turned into absorption, that is, the system entered a disc-loss
episode. 

Due to the observational gaps, it is difficult to determine the overall
time scale for the formation and dissipation of the circumstellar disc. 
The observations of \igr\ by \citet{bikmaev08} were made in Spring 2007
(low resolution) and Autumn 2007 (high-resolution) and show the \ha\ line
in absorption, although in the high-resolution spectrum the shape of the
line is reminiscent of a shell profile, i.e., with the contribution of a
small disc. If the Spring 2007 spectrum really showed an absorption profile
similar to our latest observations, then we can estimate the
formation/dissipation cycle to be about six years.

In addition to large amplitude changes in the strength of the \ha\ line,
\igr\ also display marked variations in the shape of the spectral lines. 
The most prominent spectroscopic evidence of disc activity is the long-term
V/R variability, that is, the cyclic variation of the relative intensity of
the blue ($V$) and red ($R$) peaks in the split profile of the line. The V/R
variability is believed to be caused by the gradual change of the amount of
the emitting gas approaching the observer and that receding from the
observer due to the precession of a density perturbation in the disc
\citep{kato83,okazaki91,okazaki97,papaloizou06}. Double-peak symmetric
profiles are expected when the high-density part is behind or in front of
the star, while asymmetric profiles are seen when the high-density
perturbation is on one side of the disc \citep{telting94}. 

In principle, it is possible from the data themselves to find out whether
the perturbation travels in the same direction as Keplerian orbits of the
material in the disc (prograde precession) or in opposite direction
(retrograde precession). If the motion of the density perturbation is
prograde and the disc is viewed at a high inclination angle (as in \igr),
then the $V>R$ phase should be followed by a $V=R$ phase with a strong
shell profile corresponding to the case where the perturbation lies between
the star and the observer \citep[see][for a sketch of prograde
motion]{telting94}.

In \igr, as the disc grew, the \ha\ line changed from a symmetric to an
asymmetric profile. When the disc was weak, that is, when the equivalent
width of the \ha\ line was a few Angstrom (before 2012), the intensity of
the blue and red peaks was roughly equal, $V=R$. The data reveal that the
perturbation developed between the end of 2011 and beginning of 2012. From
August 2012, a blue dominated profile is clearly present ($V>>R$). However,
the V/R ratio showed a fast decrease through the 2012 observations
(Fig.~\ref{specpar}), indicating that the V$>$R phase was coming to an end.
This was confirmed by the January 2013 observations, where an extreme
red-dominated profile is seen, $V/R\approx -0.9$.

It is worth noticing the extremely fast V/R time scales.  Although it is
not possible to pin down the exact moment of the onset of the V/R cycle due
to an observational gap of about a year (September 2011-August 2012), the
changes occurred very rapidly once the cycle started. During the $V>R$
phase, the V/R ratio changed from $\sim+0.4$ to $\sim+0.1$ in less than one
month, September-October 2012 (see Fig.~\ref{haprof} and Table~\ref{red}).
Likewise, the change from a blue-dominated $V>R$ to a red-dominated $V<R$
profile occurred in just two months (October-December 2012). If the motion
is prograde, a $V=R$ phase with a strong shell profile should be observed
in between the blue-dominated and red-dominate phases. We seem to have
missed most of this phase, which would have occurred between October and
December 2012. That is, in just two months the density perturbation must
have gone through the shell $V=R$ phase and most of the $V<R$ phase.  
Extrapolating this behaviour, we estimate the duration of a whole
revolution to be of 6--9 months.  These changes are among the fastest in a
BeXB.  The very short time scale of the observed V/R variations 
rises the question of whether these spectral changes are modulated by the
orbital period. Phase-locked V/R variations have been observed for various
Be binaries with hot companions \citep[][and references
therein]{harmanec01}, but possibly only on one BeXB (4U\,1258--61).
\citet{corbetgx86} found that the probability that the V/R variability
observed in 4U\,1258--61 was modulated by the X-ray flare period of $\sim
132$ d, which was proposed to be the orbital period of the system, was
$\sim$87\%.  Table~\ref{comp} gives a few well-studied characteristic time
scales of BeXBs: the orbital period, $P_{\rm orb}$, the V/R quasi-periods,
$T_{\rm V/R}$, and the approximate duration of the formation/dissipation of
the disc, $T_{\rm disc}$.

In common with other BeXBs \citep{reig05b,reig10b},
asymmetric profiles are not seen until the disc reaches certain size and
density. During the initial stages of disc growth, the shape of the \ha\
line is always symmetric. Only when \ew\ $\simmore -6$ \AA\ the \ha\ line
displays an asymmetric profile. \igr\ also agrees with this
result. As can be seen in Fig.~\ref{haprof}, all the spectra during the
period 2009--2011 show a symmetric profile and a small \ew. The star took
all this time to build the disc. Once a critical size and density was
reached (some time at the beginning of 2012),  the density perturbation 
developed and started to travel around in the disc. This result is also
apparent from a comparison of the two panels in Fig.~\ref{specpar}. Before
MJD 55800, $V\approx R$ and \ew\ low. After MJD 56000, a strong asymmetric
emission profile is seen.

Although the maximum \ew\ measured is relatively small, \ew $\approx -8$
\AA, compared to other BeXB, it is consistent with a well developed disc. 
In systems viewed edge-on, the maximum \ew\ is much smaller than in the
face-on case, because the projected area of the optically thick disc on the
sky is much smaller \citep{hummel94,sigut13}.  On the other hand, although
a fully developed disc must have been formed, it may not extend too far
away from the star. The fast V/R changes favoured a relatively compact
disc, where the density perturbation is capable to achieve a complete
revolution in a few months.

\subsection{Shell lines and disc contribution}

The Be star companion in \igr\ is a shell star as implied by the deep
central absorption between the two peaks of the \ha. This central
depression clearly goes beyond the continuum. The shell profiles are
thought to arise when the observer's line of sight toward the central star
intersects  parts of the disc, which is cooler than the stellar photosphere
\citep{hanuschik95,rivinius06}.   Statistical studies on the distribution
of rotational velocities of Be stars are consistent with the idea that Be
shell stars are simply normal Be stars seen near edge-on, that is, seen at
a large inclination angle \citep{porter96}. Our observations agree with
this idea. Figure~\ref{widthcomp} clearly shows that the spectral lines are
significantly narrower and deeper when the disc is present. The narrower
lines would result from the fact that the disc conceals the equator of the
star, where the contribution to the rotational velocity is largest. The
deeper lines would result from absorption of the photospheric emission by
the disc. Both circumstances require a high inclination angle. Further
evidence for a high inclination angle is provided by the photometric
observations. Both, positive and negative correlations between the 
emission-line strength and light variations have been observed and
attributed to  geometrical effects \citep{harmanec83,harmanec00}.  Stars
viewed at very high inclination angle show the inverse correlation because
the inner parts of the Be envelope  block partly the stellar photosphere,
while the small projected area of the disc on the sky keeps the disc
emission to a minimum. In stars seen at certain inclination angle,
$i\simmore i_{\rm crit}$, the effect of the disc is to increase the
effective radius of the star, that is, as the disc grows an overall (star
plus disc) increase in brightness is expected. The value of the critical
inclination angle is not known but a rough estimate based on available data
suggest $i_{\rm crit}\sim 75^{\circ}$ \citep{sigut13}. \igr\ exhibits the
inverse correlation, that is, it becomes fainter at the beginning
of a new emission episode ($\sim$ JD 2,455,800, see Fig.~\ref{specpar}). 

We have assesed the contribution of the disc on the width of helium lines,
which is the main parameter to estimate the star's rotational velocity, by
measuring the FWHM of the \ion{He}{I} 6678 \AA\ over time and by
determining the rotational velocity when the disc was present. A difference
of up to 7 \AA\ was measured between the witdh of the \ion{He}{I} 6678 \AA\
line with and without disc (see Fig.\ref{hei}). Repeating the calculation
performed in Sect.~\ref{rotvel} on the \ion{He}{I} 4026 \AA, 4387 \AA, and
4471 \AA\ when the disc was present and using the WHT December 2012
spectra, we obtain  $v\sin i=170\pm20$ km s$^{-1}$. Thus, the contribution
of the disc clearly underestimates the true rotational velocity.

In \igr, the shell profile seems to be a permanent feature.  We do not
observe any transition from a shell absorption profile to a Be "ordinary"
(i.e., pure emission) profile. In models that favour geometrically thin
discs with small opening angles, this result implies that the inclination
angle must be well above  $70^{\circ}$ \citep{hanuschik96a}. Thus, with
such a large inclination angle, the true rotational velocity is estimated
to be $v_{\rm rot}\sim$380-400 km s$^{-1}$, and the ratio of the equatorial
rotational velocity over the critical break-up velocity, $w=v_{\rm
rot}/v_{\rm crit}\sim0.8$\footnote{The break-up velocity of a B1Ve star
is $\sim 500$ km s$^{-1}$ \citep{porter96,townsend04,cranmer05}.}

If gravity darkening is taken into account, then the fractional rotational
velocity would be even larger. Gravity darkening results from fast
rotation. Rapidly rotating B stars have centrifugally distorted shapes with
the equatorial radius larger than the polar radius. As a result, the poles
have a higher surface gravity, and thus higher temperature. Gravity
darkening breaks the linear relationship between the line width and the
projected rotational velocity and makes fast rotators display narrower
profiles, hence underestimate the true rotational velocity. The reduction
of the measured rotational velocities with respect to the true critical
velocity amounts to 10--30\%, with the larger values corresponding to the
later spectral subtypes \citep{townsend04}. Correcting for gravity
darkening, the rotational velocity would be $v_{\rm rot} \sim \approx 450$
km s$^{-1}$ (assuming $i=80^{\circ}$) and the fractional rotational
velocity of the Be companion of \igr\  $w \approx0.9$, confirming the idea
that shell stars are Be stars rotating at near-critical rotation limit. We
caution the reader that the values of the break-up velocity assume that it
is possible to assign to each Be star a mass and radius equal to that of a
much less rapidly rotating B star, e.g., from well studied eclipsing
binaries. Given that there is not a single direct measurement of the mass
and radius for any known Be star, the break-up velocity should be taken as
an approximation \citep[see e.g][]{harmanec00}.

\section{Conclusion}

We have performed optical photometric and spectroscopic observations of the
optical counterpart to \igr. Our observations show that \igr\ is a
high-mass X-ray binary with a Be shell type companion.  Its long-term
optical spectroscopic variability is characterised by global changes in the
structure of the equatorial disc. These global changes manifest
observationally as asymmetric profiles and significant intensity
variability of the \ha\ line. The changes in the strength of the line are
associated with the formation and dissipation of the circumstellar disc. At
least since 2009, \igr\ has been in an active Be phase that ended in mid
2013, when a disc-loss episode was observed. The entire
formation/dissipation cycle is estimated to be six years, although given
the lack of data before 2009, this figure needs to be confirmed by future
observations. In contrast, the  V/R variability is among the fastest
observed in Be/X-ray binaries with characteristic timescales of the order
of few weeks for each V/R phase.  

The absence of the disc left the underlying B star exposed, allowing us to
derive its astrophysical parameters. From the ratios of various metallic
lines we have derived a spectral type B1IVe. The width of He I lines imply
a rotational velocity of $\sim370$ km s$^{-1}$. Using the photometric
magnitudes and colours we have estimated the interstellar colour excess
$E(B-V)\sim0.76$ mag, and the distance $d\sim$ 8.5 kpc.

The presence of shell absorption lines indicate that the line of sight to
the star lies nearly perpendicular to its rotation axis. Although the
Balmer lines show the most clearly marked shell variability, the helium lines
are also strongly affected by disc emission, making them narrower than in
the absence of the disc.

\begin{acknowledgements}

We thank the referee P. Harmanec for his useful comments and suggestions
which has improved the clarity of this paper. We also thank observers P.
Berlind and M. Calkins for performing the FLWO observations and I.
Psaridaki for helping with the Skinakas observations.  Skinakas Observatory
is a collaborative project of the University of Crete, the Foundation for
Research and Technology-Hellas and the Max-Planck-Institut f\"ur
Extraterrestrische Physik. The WHT and its service programme (service
proposal references SW2012b14 and SW2013a19) are operated on the island of
La Palma by the Isaac Newton Group in the Spanish Observatorio del Roque de
los Muchachos of the Instituto de Astrof\'{\i}sica de Canarias. This paper
uses data products produced by the OIR Telescope Data  Center, supported by
the Smithsonian Astrophysical Observatory. This work has made use of NASA's
Astrophysics Data System Bibliographic Services and of the SIMBAD database,
operated at the CDS, Strasbourg, France.

\end{acknowledgements}

\bibliographystyle{aa}
\bibliography{artBex_bib}

\end{document}